\documentclass[twocolumn, showpacs, aps, prl, superscriptaddress, floatfix]{revtex4}
\usepackage{graphicx}
\usepackage{amsmath}
\begin{document}

\title{Bayesian Error Estimation in Density Functional Theory}
\author{J. J. Mortensen}
\author{K. Kaasbjerg}
\author{S. L. Frederiksen}
\author{J. K. N{\o}rskov}
\affiliation{CAMP and Department of Physics, Technical University of
  Denmark, DK-2800 Lyngby, Denmark}
\author{J. P. Sethna}
\affiliation{Laboratory of Atomic and Solid State Physics (LASSP), Clark
  Hall, Cornell University, Ithaca, NY 14853-2501, USA}
\author{K. W. Jacobsen}
\affiliation{CAMP and Department of Physics, Technical University of
  Denmark, DK-2800 Lyngby, Denmark}
\date{\today}

\begin{abstract}
  We present a practical scheme for performing error estimates for
  Density Functional Theory calculations. The approach which is based
  on ideas from Bayesian statistics involves creating an ensemble of
  exchange-correlation functionals by comparing with an experimental
  database of binding energies for molecules and solids. Fluctuations
  within the ensemble can then be used to estimate errors relative to
  experiment on calculated quantities like binding energies, bond
  lengths, and vibrational frequencies. It is demonstrated that the
  error bars on energy differences may vary by orders of magnitude for
  different systems in good agreement with existing experience.
\end{abstract}


\maketitle

Over the past few decades the use of Density Functional Theory
(DFT)\cite{Hoh64} to predict structures, energetics, and other
properties of atomic-scale systems has spread to many different fields
and the number of applications has grown enormously. Today
applications may vary from studies of chemical reactions in
heterogeneous catalysis\cite{Hon05} through geophysical investigations of
melting at the physical conditions of the Earth's
core\cite{Lai00,Alf99} to studies of biomolecular systems like
DNA\cite{Sky05, Art03}. The general usefulness of the calculations lies in
their unbiased ``first principles'' character and the relatively high
degree of predictive power and reliability which has been established.
With respect to the latter, it is however often difficult to assess
directly to which extent a calculated quantity -- this being a
molecular bond length or some other property -- is to be trusted. In
practice the evaluation often falls back exclusively on the experience
and acquired insight of the person performing the calculation.

In this letter we present a scheme for systematic error evaluation
within the generalized-gradient approximation (GGA) DFT. The scheme is
based on ideas from Bayesian statistics\cite{Jay03} in which an
ensemble of models or model parameters are assigned probabilities by
comparing to a database of experimental results. The ensemble
generated can then subsequently be used to estimate error bars on
model predictions within the considered class of models. The scheme is
simple to apply and amounts in the end to only a few additional
non-self-consistent evaluations of exchange-correlation functionals.
In spirit, it is in fact close to a rather common practice within the
field of DFT-GGA calculations: To asses the validity of a calculated
DFT-GGA result it is common to try out different versions of the
GGA-functional or perhaps to compare with a
local-density approximation (LDA) result. The scheme presented here
provides a systematic framework for such an approach.

The statistical approach we use is inspired by Bayesian statistics
\cite{Jay03} and was further developed in the context of modeling
complex biomedical networks \cite{Bro03} and for construction of
interatomic potentials \cite{Fre04}. The main ingredients in the
approach are a model $M$ which is given by a set of parameters
$\theta$, and a database $D$. In our case the model will be a GGA-type
exchange-correlation functional described through a number of
parameters $\theta$ and the database $D$ will consist of
experimental atomization/cohesive energies $E_k^{\text{exp}}$ for a
collection of molecules and solids (details below). We now define a
probability distribution in the space of model parameters (given the
choice of model and the database) through
\begin{equation}
  \label{eq:prob}
  P(\theta|MD) \sim  \exp (-C(\theta)/T),
\end{equation}
where $C$ denotes a standard least-squares cost function $C(\theta) =
\frac{1}{2} \sum_k (E_k(\theta) - E^{\text{exp}}_k)^2$ with
$E_k(\theta)$ being the atomization/cohesive energy of system $k$ in
the database calculated with the parameters $\theta$. The ``effective
temperature'' $T$ determines the spread of the ensemble. In simple
fitting procedures only the best-fit parameters $\theta_\text{b.f.}$, which are
obtained when the cost function takes on its minimal value $C_\text{b.f.}$, are
considered. Here, in contrast, a whole ensemble of parameter sets are
considered leading to a spread of values on model predictions.
Following Ref.~\onlinecite{Fre04}, we take the value of the effective
temperature to be given by the minimal (best fit) value of the cost
function $C_\text{b.f.}$ as $T=2 C_\text{b.f.}/N_\text{p}$, where $N_\text{p}$ is the
number of parameters. For a harmonic cost function each parameter will
then on the average contribute an additional cost of $T/2 =
C_\text{b.f.}/N_\text{p}$ so that the ensemble will probe model parameters where
the cost function is in the range from the minimal value $C_\text{b.f.}$ up to
of the order a few times this value. This choice was demonstrated to
work well in the case of error estimation for interatomic potentials
\cite{Fre04}.

\begin{figure}[htbp]
  \centering
  \includegraphics[width=\linewidth,clip=]{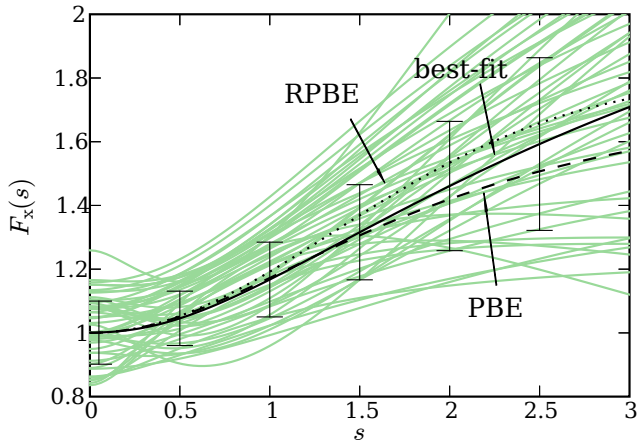}
  \caption{Exchange enhancement factors as a function of dimensionless
    electron density gradient.  The gray lines show enhancement
    factors drawn from the ensemble $\exp(-C(\theta) / T)$.
    The dashed, dotted and full lines show enhancement factors for
    PBE, RPBE and the best fit respectively.}
  \label{fig:enhancement}
\end{figure}

{\em The model} we shall consider is GGA-DFT \cite{Lan80} where the
exchange functional is a local function of the density $n$ and its
dimensionless gradient $s=|\nabla n|/(2k_\text{F}n)$
($n=k_\text{F}^3/(3\pi^2)$).  Several suggestions for different
mathematical forms of the exchange-functional within GGA
exist\cite{Lan80, Boe00, Per96}. A
commonly used class of these including PW-91\cite{Per92},
PBE\cite{Per96}, revPBE\cite{Zha98} and RPBE\cite{Ham99} differ by
only the choice of the so-called enhancement factor $F_\text{x}(s)$ in
the exchange energy $E_\text{x}$:
\begin{equation}
E_{\text{x}}[n] =
\int d\mathbf{r}\, n(\mathbf{r}) \epsilon_{\text{x}}^{\text{LDA}}(n(\mathbf{r}))
F_{\text{x}}(s),
\end{equation}
where $\epsilon_{\text{x}}^{\text{LDA}}(n)=-3k_\text{F}/(4\pi)$ (for a
spin polarized density we have $E_{\text{x}}[n_\uparrow, n_\downarrow]
= (E_{\text{x}}[2n_\uparrow] + E_{\text{x}}[2n_\downarrow]) / 2$).
The enhancement factors for PBE and RPBE are shown in
Fig.~\ref{fig:enhancement}.  In the following we shall expand the
enhancement factor as
\begin{equation}
  \label{eq:enhancement}
F_{\text{x}}(s) = \sum_{i=1}^{N_\text{p}} \theta_i \left ( \frac{s}{1 +
    s}\right )^{2i-2},
\end{equation}
regarding the $\theta$'s as free parameters. We use three parameters
($N_\text{p}=3$) which a train/test check for our database has shown
to give the optimal fit without over-fitting.
The model space could be extended in future work to include a fraction
of exact exchange as for the B3LYP functional\cite{Bec93}.

{\em The database} we use consists of the experimental atomization
energies of the molecules H$_2$, LiH, CH$_4$, NH$_3$, OH, H$_2$O, HF,
Li$_2$, LiF, Be$_2$, C$_2$H$_2$, C$_2$H$_4$, HCN, CO, N$_2$, NO,
O$_2$, F$_2$, P$_2$ and Cl$_2$ and the experimental cohesive energies
(per atom) of the solids Na, Li, Si, C, SiC, AlP, MgO, NaCl, LiF, Cu
and Pt. In the cost function all systems in the database appear with
the same unit weight.

All calculations are performed with a real-space multigrid DFT
code\cite{Mor05} using the projector augmented wave method\cite{Blo03}
to describe the core regions.  All calculated energy differences have
been converged to an accuracy better than 50 meV with respect to
number of grid points, unit-cell size (for the molecules and atoms)
and number of k-points (for the solids).  The electron density is
calculated self-consistently using the PBE-functional and the
evaluation of the exchange-correlation energy for other enhancement
factors are performed using the PBE density.  This is a very good
approximation due to the variational principle. Energies are
calculated at the calculated equilibrium bond distances.

Since Eq.~(\ref{eq:enhancement}) is linear in the parameters $\theta$,
the total energy of a given system will also be
linear in $\theta$:
\begin{equation}
\label{eq:linear}
E(\theta) = E_0 + \sum_{i=1}^3 \Delta E_i \theta_i,
\end{equation}
where the coefficients $E_0$ and $\Delta E_i$ only have to be
calculated once for each system. It is therefore very fast to
calculate energy values for different enhancement factors in the
ensemble.

Minimizing the cost function with respect to the three parameters
leads to the best-fit enhancement factor shown in
Fig.~\ref{fig:enhancement} corresponding to the parameters
$\theta_\text{b.f.} = (1.0008, 0.1926, 1.8962)$.  The function is seen
to follow quite closely the PBE enhancement factor at low values of
the gradient $s$.  In particular it is nearly one in the homogeneous
limit ($s=0$) which is exclusively a result of the fitting. For
$s$-values greater than 1.5 the best-fit enhancement factor increases
more steeply than PBE being more similar to the RPBE factor. In
Table~\ref{tab:errors} the resulting errors are shown for LDA, PBE,
RPBE and for the best fit.  RPBE performs better on the molecules and
PBE is better for the solids; the best fit represents a compromise
between the two.  We would like to stress that the main point of this
letter is not to derive an improved functional.  Much experience has
been acquired concerning how well different GGA functionals work for
different systems \cite{Kur99, Sta03, Sta04} and we do not expect to
obtain a large overall improvement within this simple GGA framework.
But as we shall see in the following the ensemble construction allows
for realistic evaluation of the error bars on calculated quantities.

\begin{table}[htbp]
  \caption{Errors in DFT atomization energies and cohesive energies (in eV)
    relative to experiment.  All calculations are based on
    self-consistent PBE densities.  Experimental
    numbers are taken from Refs.~\onlinecite{Per96, Sta04, Kit96,
      Lee97, Pau96, Day03, Sch98}.}
\begin{tabular}{lcccc}
  \hline
  \hline
  error     & LDA               & PBE          & RPBE          & best fit \\
  \hline
  molecules: & & & & \\
  \hline
  mean abs. &  1.46             &  0.35        &  0.21         &  0.24 \\
  mean      &  1.38             &  0.28        & -0.01         &  0.12 \\
  max. (-)  & -0.35(H$_2$)      & -0.22(H$_2$) & -0.32(CH$_4$) & -0.26(Li$_2$)\\
  max. (+)  &  3.07(C$_2$H$_4$) &  0.89(O$_2$) &  0.46(O$_2$)  &  0.71(O$_2$) \\
  \hline
  solids: & & & & \\
  \hline
  mean abs. &  1.35             &  0.16        &  0.40         &  0.27 \\
  mean      &  1.35             & -0.09        & -0.40         & -0.24 \\
  max. (-)  &                   & -0.72(Pt)    & -1.37(Pt)     & -0.94(Pt) \\
  max. (+)  &  2.73(Pt)         &  0.36(C)     &               &  0.15(C) \\
  \hline
  all: & & & & \\
  \hline
  mean abs. &  1.42             &  0.28        &  0.28         &  0.25 \\
  mean      &  1.37             &  0.14        & -0.16         & -0.02 \\
  \hline
  \hline
\end{tabular}
\label{tab:errors}
\end{table}

The cost function appearing in the probability distribution
Eq.~(\ref{eq:prob}) is very nearly quadratic in the model parameters in
the relevant range of parameter space. We can
therefore expand the exponent in the probability distribution as
$C(\theta)/T = \text{const.} + \frac{1}{2}\Delta\theta^{T} A
\Delta\theta$, where $A$ is a symmetric matrix.
With $U$ being the unitary matrix that diagonalizes $A$
($AU=U\Lambda$), we can finally write the parameters of the
enhancement factors in the ensemble as
\begin{eqnarray}
  \label{eq:parameters}
\theta & = & \theta_\text{b.f.} + U\Lambda^{-1/2}\alpha \nonumber\\
& = & \theta_\text{b.f.} + 
\begin{pmatrix}
   0.066 & 0.055 & -0.034 \\
  -0.812 & 0.206 &  0.007 \\
   1.996 & 0.082 &  0.004
\end{pmatrix}
\begin{pmatrix}
  \alpha_1 \\
  \alpha_2\\
  \alpha_3
\end{pmatrix},
\end{eqnarray}
where the $\alpha_1$, $\alpha_2$, and $\alpha_3$ are stochastic
variables which are Gaussian distributed with unit width:
$\mathcal{P}(\alpha_i) \sim \exp(-\alpha_i^2/2)$. Using this
formula it is very easy to generate a properly distributed ensemble of
enhancement factors as shown in Fig.~\ref{fig:enhancement}.

The key suggestion of this letter is that for a given calculated
observable, say a bond length of a molecule, the variation of the
calculated value of this observable within the ensemble of enhancement
factors provides a useful estimate of how large the error of the best
fit value is compared to experiment.  From an ensemble of parameters,
$\theta^1, \theta^2, \ldots, \theta^N$, generated from
Eq.~(\ref{eq:parameters}), the standard deviation
$\sigma_\text{BEE}(O)$ which we shall refer to as the Bayesian error
estimate (BEE) of the observable $O$ can be determined. Considering
$O$ as a function of $\theta$ the BEE is evaluated as
\begin{equation}
  \label{eq:bee}
  \sigma_\text{BEE}(O) = \sqrt{\frac{1}{N} \sum_{\mu=1}^N
    \left ( O(\theta^\mu)- O_\text{best-fit} \right )^2 },
\end{equation}
In the simple case where the observable is approximately linear in the
parameters $\theta$ the BEE can be calculated without explicitly
generating an ensemble through $\sigma_\text{BEE}(O) = (
\sum_{i=1}^{3} \left(
  \partial O / \partial \alpha_i \right)^2)^{1/2}$ and
Eq.~(\ref{eq:parameters}).  Further details can be found in our
implementation of the approach in the Atomistic Simulation
Environment\cite{ASE}.

The ensemble in Eq.~(\ref{eq:bee}) is around the best-fit enhancement
factor corresponding to $\alpha_i=0$. However, considering that the
ensemble of enhancement factors (Fig.~\ref{fig:enhancement}) is quite
wide compared with the difference between for example the PBE and the
best-fit functional it seems reasonable to alternatively apply the
fluctuations around either the PBE or RPBE functional.

It is useful to consider the ratio $(O_\text{best-fit} -
O_\text{exp})/ \sigma_\text{BEE}(O)$ of the actual error relative to
the estimate. For any given observable in a particular system this
ratio is just a single number so in order to assess whether our
approach produces reliable error estimates from a statistical point of
view we need to look at the distribution of ratios for several
observables and systems.

\begin{figure}[htbp]
  \centering
  \includegraphics[width=\linewidth,clip=]{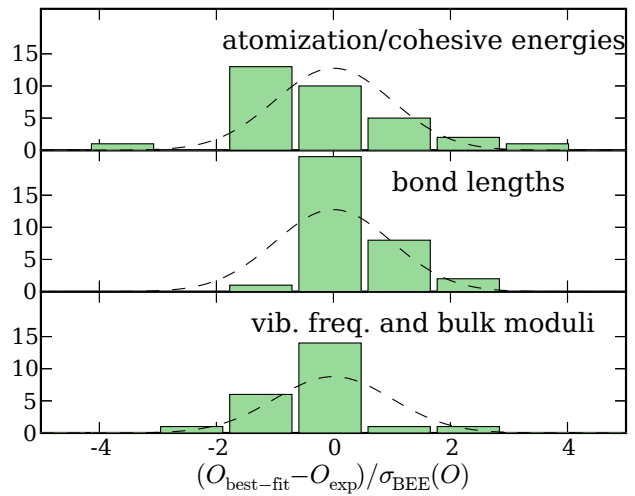}
  \caption{Histograms of actual error relative to the BEE
    $(O_\text{best-fit} - O_\text{exp}) / \sigma_\text{BEE}(O)$ for
    different quantities.  Top: atomization/cohesive energies.
    Middle: bond lengths and lattice constants.  Bottom: vibrational
    frequencies and bulk moduli.  The dashed lines show the expected
    normal distribution.  Experimental numbers are taken from
    Refs.~\cite{Joh93, Sta04, Rod92, Sch98, Kur99}.}
  \label{fig:ae}
\end{figure}

In Fig.~\ref{fig:ae} we show histograms of the relative error for
different observables for all the (molecular and solid) systems in the
database. The upper panel shows the distribution in the case of the
atomization/cohesive energies.  As can be seen the distribution agrees
quite well with a Gaussian distribution of unit width indicating that
the error estimates are in fact reasonable for the binding energies.
The individual standard deviations for the cohesive energies are in
the range from 0.09 eV for Na to 0.75 eV for Al and for the
atomization energies the range is from 0.07 eV for Li$_2$ to 0.60 eV
for C$_2$H$_4$.  As an example, the GGA estimate of the cohesive
energy of Na (experimental value is 1.11 eV) is $1.02\pm0.09$ eV.  The
middle panel in Fig.~\ref{fig:ae} shows the relative error histogram
for the bond lengths (for both molecules and solids) and relative
errors for the molecular frequencies and solid bulk moduli is shown in
the lower panel\cite{footnote2}.  For both distributions, we see that
the BEE's give reasonable estimates of the actual errors.

\begin{figure}[htbp]
  \centering
  \includegraphics[width=\linewidth,clip=]{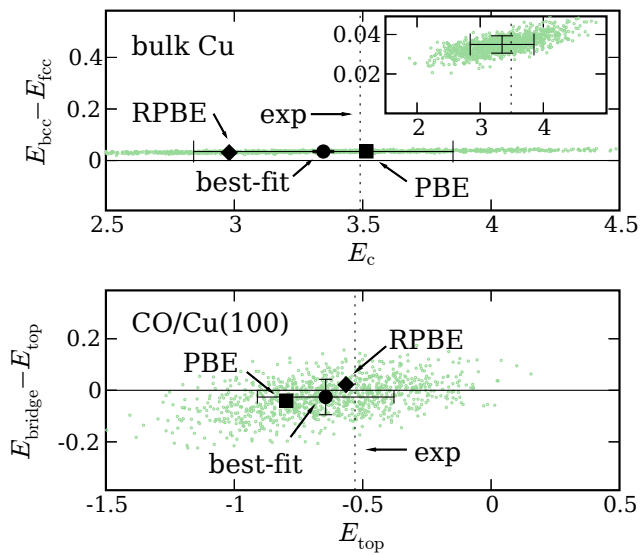}
  \caption{Upper panel: Calculated ensemble for cohesive energy
    (x-axis) and bcc-fcc energy difference (y-axis) for a copper
    crystal. The BEE's are indicated by error bars. The inset uses
    rescaled axes. Lower panel: Calculated ensemble for binding energy
    (x-axis) and bridge-top energy difference for CO on a Cu(100)
    surface. Values for the experimentally preferred
    states\cite{Kit96, Vol01} (fcc and top) are indicated by vertical
    dotted lines. Units: eV.}
  \label{fig:Cu}
\end{figure}

It is well-known for experienced users of DFT calculations that the
reliability with which energy differences can be calculated vary
dramatically depending on the particular system. The BEE catches this
behavior as can be seen for example by comparing the cohesive energy
and the bcc-fcc structural energy difference for bulk copper
(Fig.~\ref{fig:Cu}). The BEE for the cohesive energy is 0.5 eV while
the error bar on the structural energy difference is two orders of
magnitudes smaller (4 meV).  The small structural energy difference is
seen to be significantly greater than zero.  The high reliability with
which small structural energy differences can be calculated for bulk
metals is confirmed by the fact that for almost all metals the correct
equilibrium structures are predicted from the
calculations\cite{Skr85}.

For chemisorption systems the BEE for the energy difference between
chemisorption at two difference surface sites may also be somewhat
smaller than the error bar for the total chemisorption energy as
illustrated in the case of CO on Cu(100) in Fig.~\ref{fig:Cu}. However
as can be seen from the figure the error bar on the site-preference is
so large that the preferred chemisorption site cannot be reliably
determined. This is in good agreement with the fact that for a number
of CO-metal chemisorption systems DFT-GGA calculations do in fact
predict a wrong chemisorption site\cite{Fei01}.

It should be noted that in general the error estimates depend on both
the choice of database and the class of models (here the GGA's). This
means that if some piece of physics is not present in the database or
if the model is completely unable to describe a particular feature
unrealistic estimates may occur. 
For example all GGA's are unable to properly describe the long-range
van der Waals interactions and hence the BEE's will be unrealistic for
that type of interactions.

We further note, that it may be possible to reduce the error bars in
some cases by picking a database focused on a particular type of
systems. Using for example only the subset of molecular systems in our
database a different best-fit functional is obtained where the mean
absolute error is only 0.15 eV compared to the 0.24 eV in Table
\ref{tab:errors}. With this functional the fluctuations and hence the
BEEs will be reduced by almost a factor of two. However, applying this
functional to bulk metals would lead to larger errors which would be
underestimated.

Summarizing, we propose a simple way to estimate a large portion of
the systematic error for DFT-GGA calculations, requiring only a few
extra non-self-consistent energy calculations to calculate the error
bars of any observable which is a function of energies.

We acknowledge support from the Carlsberg Foundation and the Danish
Center for Scientific Computing through Grant No.\ HDW-1101-05.


\begin{thebibliography}{10}
\bibitem{footnote2}
Vibrational frequencies are only calculated for a subset of the molecules.
\end{thebibliography}
\end{document}